\documentclass[letterpaper, 10pt, conference]{ieeeconf}
\IEEEoverridecommandlockouts 
\usepackage{amsmath,amscd,amssymb,amsgen,amsfonts,amsbsy}
\usepackage{mathrsfs}
\usepackage[utf8x]{inputenc}
\usepackage{color}
\usepackage{textcomp}
\usepackage{float}
\usepackage{latexsym,graphicx}

\usepackage{tikz}
\usetikzlibrary{automata,arrows,positioning,calc}
\usepackage{url}
\usepackage{cite}
\usepackage{algorithmic}
\usepackage[ruled,lined]{algorithm2e}

\newcommand{\remove}[1]{}
\newcommand{\rahul}[1]{\textbf{RAHUL}: \textcolor{blue}{#1}}
\newcommand{\comments}[1]{}

\renewcommand{\baselinestretch}{1}

\newtheorem{proposition}{Proposition}

\newtheorem{definition}{Definition}

\makeatother

\title{Online repeated posted price auctions with a demand side platform}

\author{
 \begin{tabular}{ccc}
   Rahul Meshram & & Kesav Kaza \\
    Deptt. of Elecl. Engg.    
    & & Deptt. of Elecl. Engg. \\
    IIT Madras, Chennai INDIA 
     &  &   IIT Bombay, Mumbai INDIA   
  \end{tabular}
}

\begin{document}

\maketitle

\begin{abstract}
  We consider an online ad network problem in which an ad exchange auctions ad slots   
  and intermediaries called demand side platforms (DSPs) buy these ad slots for their clients (advertisers). 
  An intermediary represents multiple advertisers. Different types of ad slots 
  are auctioned by the ad exchange, e.g., video ad, banner ad etc. We study 
  repeated posted price auctions for homogeneous and  heterogeneous items when there is an intermediary. 
  In a posted price auction, the auctioneer sets a fixed reserve 
  price. The buyer can accept the price and win the ad slot or reject the price.

  We analyze the system from the auctioneer's perspective and show that the optimal reserve price is 
  dynamic for heterogeneous items. We also investigate system from intermediary's
  perspective and  devise algorithms for  scheduling advertisers.  
  Often the advertisers have budget constraints and impression constraints. We formulate a revenue optimization problem  at the 
  intermediary and also consider the problem of scheduling advertisers with budget and impression constraints. Finally, we present a numerical study  
 for the single seller and advertiser model which considers various valuation distributions such as uniform, exponential and lognormal.
\end{abstract}

\section{Introduction}


In the age of information and internet, users often visit to different webpages for various goal, for example news (www.thehindu.com), travel  (www.irctc.co.in), etc. The webpages generate revenue from having ad display slots on their page for advertisement. With increasing use of internet, mobile internet and e-commerce platforms, the online ad display market has grown significantly in last few years. The total revenue generated from internet advertising in US was $ \$ 107.5 $ billion for $2018,$ \cite{US2019}. The digital advertising market in India  was around $ \$1.3$ billion for $2018,$ \cite{India2019}. This is expected to grow further in coming years. Motivated from these developments, in this paper we study  an auction mechanism for the online ad display market.

Various online content publishers have a large number of webpages. The objective of these publishers is to maximize their revenue via displaying ads. Publishers can enter into contracts with advertisers to display ads in ad slots on available on their webpages.
Another way for publishers to optimize their revenue from ads is by selling ad slots using an auction mechanism, instead of fixed contracts with advertisers. These  auctions are conducted at Ad exchanges. Examples of ad exchanges include  Google’s Double Click, OpenX, Yahoo!’s Right Media, etc.

\begin{figure}
  \includegraphics[scale= 0.43]{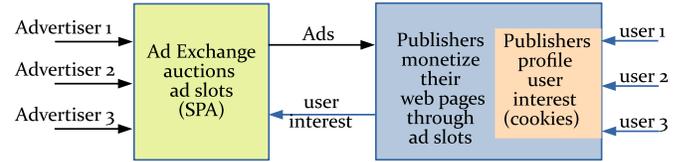}
 	  \caption{Online ad auction  model }
	\label{Fig1:Ad-auction-Model}
\end{figure}

We now briefly describe how online ad auctions work. Consider a users arriving at a webpages. The publisher approaches an ad exchange with ad display slots. Different advertisers participate in an ad-slot auction held by the ad exchange. The  publisher shares user information (from cookies) with advertisers via the ad exchange. The advertisers that match the user's interests, participate in the auction and bid (maximum willingness to pay) for the ad slot. This bid price depends on their valuation for the ad slot. The auctioneer runs the second price auction (SPA) mechanism, where ad slot is allocated to advertiser with highest bid price and the winner pays second highest bid price. This auction is performed in a  timescale of milliseconds. Each ad slot is referred to, as an impression.  Online ad auction model is illustrated in Fig.~\ref{Fig1:Ad-auction-Model}.

Large auctioneers (ad exchanges) may sell billions of impressions in a day using the SPA mechanism. When the user clicks on an ad slot, (s)he is sent to the advertiser's webpage. The advertiser's valuation for the ad slot depends on user's click-through probability.
   The goal of the auctioneer is to maximize its long term revenue thorough optimal reserve pricing. Whereas the objective of advertisers is to determine the bid price for each auction such that their  long term revenue is maximized. In \cite{Muthukrishnanan09}, different models of ad exchanges are discussed.


%
%
%
%
%

Often advertisers   participate in auctions through intermediaries called as demand side platforms (DSPs). This may be because most of the advertisers have low budget and might incur additional cost for maintaining their services and designing optimal bid strategies. Instead, a DSP represents the multiple advertisers in online ad auctions. These advertisers can enter into contracts with DSP for buying impressions and displaying their ads. Advertisers may have targeting constraints, budget constraints or impression constraints. Sometimes, advertisers might prefer  some types of ads over others. For example, one may have preference for video ads over banner ads. 
Further, advertisers may have different willingness to pay for different types of ads. A DSP can have few hundreds of advertisers. \textit{ A DSP needs a policy to  schedule these advertisers in online ad auctions, and, a bidding algorithm that  maximizes its total utility while meeting the requirements of the advertisers.} In online ad auctions, there can be many DSPs participating in auctions. This may influence the bid prices in auctions and affect expected revenue of the auctioneer and the publishers.

In this paper we consider a single seller and buyer (DSP or a advertiser) model. This is motivated from the ad auction problem where an advertiser or a DSP participates in large number of auctions. The SPA mechanism with a single buyer and  seller is called as posted price auction (PPA) mechanism. In this mechanism, seller sets a reserve price on an item. The  buyer accepts the offer if the reserve price set by auctioneer is lower than its own valuation, otherwise it rejects the offer. Repeated auctions with posted price mechanism are referred to repeated posted price  auctions. This is a simplified model of SPA mechanism. 



\subsection{Related work}
We now discuss some relevant work on auctions, online ad auctions and auction with intermediaries. A second price auction mechanism is first introduced by Vickery in \cite{Vickery61}, this is also referred to as Vickery auction. 
In SPA, there is a seller and set of potential buyers. Each buyer values a good differently and is willing to pay a different amount. Each buyer knows their valuation for the good but does not know the valuations of other buyers. The seller also does not know the valuations of buyers. But the seller and buyers know that their valuation is drawn from a probability distribution function. This distribution is independent across the buyers. This model is called as \textit{ independent private value (IPV)} model. We further consider a symmetric buyer valuation in which this probability distribution is identical. 
The problem of optimal auction design for IPV model is studied in \cite{Myerson81, Laffont80, Riley81}. Here, all buyers bid for a good.  The buyer with the highest bid price wins the auction and pays the second highest bid price. In optimal auction design these authors studied equilibrium bidding behavior. It is shown that in equilibrium each buyer's dominant strategy is to report true valuation to seller. This is called as \textit{dominant strategy incentive compatible} (DSIC) auction. Further, the seller can optimize revenue by setting an optimal reserve price for SPA and it is independent of the number of buyers in SPA. For more details on auctions, see \cite{Krishna10}.

Recently, repeated SPA has been studied in \cite{Gummadi13,Balseiro15} for budget constrained online auction problem. Here, it is shown that the optimal bidding behavior of buyer in equilibrium is to shade their bid and set the bid price less than their valuation.  The buyers do not report true valuation and hence this is not DSIC auction. 
The various issues in online ad auction, and ad exchanges are mentioned in \cite{Muthukrishnanan09} and author discusses the perspectives of the ad exchange, the advertiser, the publisher, and, introduces the role of intermediaries in the ad network. The role of reserve prices in auction design using a field experiments are demonstrated in \cite{Ostrovsky09}. In \cite{Kanoria14},  dynamic reserve price based learning is introduced for unknown value distribution function, in online ad auction, and, incentive compatible constraints are also studied. A variant of posted price auction model is investigated in \cite{Celis14} and  randomized auction model is studied.

In the domain of online ad display network, there are a few works on ad auction with intermediaries (DSPs).
This study has become important due to resource constraints put in by the  advertisers while optimizing the online ad auction target criteria. In \cite{Feldman10}, the author studies auctions with intermediaries, where auctions are run at both the seller and the intermediary. For auction at the seller, the intermediaries are buyers and for auctions at intermediaries, the advertisers are buyers. And, the equilibrium behavior is  studied.  A single item auction with competing intermediary is investigated in \cite{Stavrogiannis13,Stavrogiannis14}. A few variants of auctions with intermediaries are studied in \cite{Balseiro17, Balseiro18} where the authors discuss dynamic bidding strategies for budget constrained advertisers.
 Dynamic pricing model for real time bidding in online ad display is studied in \cite{Chen14}. Another direction of research on auctions with intermediaries is studied in \cite{Allouah17}, where authors introduce a multi-bidding concept.

None of the above models consider repeated posted price auction with an intermediary, and the problem of advertiser-scheduling faced by the intermediary. 
 


\subsection{Our contributions}

In this paper, we  study repeated posted price auction mechanisms for the auction of homogeneous  and heterogeneous items. We show that the optimal reserve price of seller is fixed all round of auctions for homogeneous items and the optimal reserve price of seller for heterogeneous item is dynamic. Later, we use this insight to study an  online ad auction system with a DSP as an intermediary. We provide optimization formulation as revenue maximization and   scheduling scheme for the advertisers at the DSP under impression constraints and budget constraints.  
We provide numerical examples for a single seller and advertiser model for uniform, exponential and lognormal distribution on valuation of advertiser. We also illustrate numerical examples with DSP model.

\section{Preliminaries and Model}
\label{sec:prelim}
Consider an online ad auction system with a seller (ad exchange) and a DSP.  
The seller has multiple units of $K$ types of items for sale and he auctions items  sequentially\footnote{Examples of different types of items include banner ad, video ad, etc. The ad exchange can observe the types of ads that can be displayed at the publisher's site.}. 
Suppose that there are $T$ rounds of auctions. We index the each round in the sequence by $t$ and $t = 1, 2, \cdots T.$ 
 Advertisers participate in auctions through  the DSP. The seller uses posted price auction mechanism. The DSP schedules an advertiser from its set of client advertisers in each round of auction by accepting the price offered by the seller. 
The seller knows the type of the item which is being autioned. The DSP also has this information ( type of item being auctioned).    

Users arrive at various websites sequentially. Different websites have different types of ads slots available for display\footnote{Example: Youtube has video ad slots as well as display banner ad slots, for the arriving user.}. Let $p_k$ be the probability that item of type $k$ is auctioned by  seller in auction round $t,$ and $\sum_{k=1}^{K} p_k =1.$ We assume that the type of items auctioned in each round are independent of the other rounds. Let $X_{k,t}$ be the indicator random variable for round $t,$ type $k,$ with  $1 \leq k \leq K$ and $1 \leq t \leq T.$  It takes value   $X_{k,t} = 1$ if item of type $k$ is auctioned in round $t;$ otherwise $X_{k,t} = 0.$   Let  $q_t$ be the price set by seller in the round $t.$ This price could depend on type of item. We shall explicitly mention this dependence of price $q_t$ on the type of item whenever necessary.

Suppose that the DSP has a set of $N$ advertisers (clients) denoted by $\mathcal{N} = \{1,2, \cdots,N\}.$ The advertisers are risk neutral.

 We assume that the seller does not know how much each advertiser is willing to pay for different types of items. Let $\tilde{v}_{i,k,t}$ be the value estimate of advertiser $i$ for type $k$ item in round $t$, i.e., the maximum amount an advertiser is willing to pay for an item.  The seller's uncertainty about the value estimate of advertiser's value is represented by a continuous probability distribution function over a finite interval. Let $F_{i,k}$ be the probability distribution function for advertiser $i'$s valuation given that type $k$ item is auctioned.  



We now discuss posted price auction mechanism. 

\subsection{Posted price auction mechanism}
In the standard posted price auction, there is a seller and a buyer. 
The buyer's value estimate $v$ is unknown to the seller, but he knows the probability distribution of the value estimate, say, $F$ over $[\underline{v}, \overline{v}]$\footnote{We need to assume that $F$ is regular; see assumption $1$ in \cite{Myerson81, Kanoria14}}. The seller sets a reserve price, say $q$ and announces it to buyer. If $v > q,$ the buyer accepts the offer, otherwise the  offer is rejected. The revenue of the seller is $q$ when the offer is accepted; it is $0$ otherwise. The payoff of the buyer is $v-q$ when the offer accepted, and zero otherwise. This strategy of the buyer is dominant strategy incentive compatible (DSIC). 
From the seller's perspective, the objective is to optimize the revenue by setting optimal reserve price. This price is the solution of following problem.
\begin{equation*}
q^{\ast} = \arg \max_{q} q\times \mathrm{Pr}{(v >q )} = \arg \max_{q} q (1 - F(q)).
\end{equation*}

\subsubsection*{Repeated posted price auction (RPPA) mechanism}
In this case there are $T$ rounds of auctions. And each round follows the PPA mechanism.
Now consider RPPA with a single buyer. 
Let $v_t$ be the valuation of the buyer for round $t;$ it is drawn from probability distribution $F$ independently for each round.  Let $q_t$ be the reserve price set by the seller in round $t.$  The buyer accepts the offer whenever $v_t \geq q_t$ and rejects it otherwise. The expected revenue of a buyer over $T$ rounds of auctions is 
\begin{eqnarray}
R_{B,PPA} = \mathbb{E} \left[\sum_{t=1}^{T} (v_t - q_t) 1_{\{q_t < v_t\}} \right].
\label{eqn:rppa-buyer}
\end{eqnarray}
The strategy of the buyer is DSIC.  The seller's expected revenue over $T$ rounds of auctions is 
\begin{eqnarray}
R_{S,PPA} = \mathbb{E} \left[\sum_{t=1}^{T} q_t 1_{\{q_t < v_t\}} \right].
\label{eqn:rppa-seller}
\end{eqnarray}
The objective of a seller is to set an optimal reserve price to maximize~\eqref{eqn:rppa-seller}. 
\subsection{DSP's problem in a repeated posted price auction}
We now consider the problem from the perspective of the DSP. It has $N$ advertisers (clients). Let $v_{k,j_t}$ be the valuation of advertiser $j_t =i$ for item of type $k$ in auction $t.$  Suppose that $q_{k,t}$ be the price set by seller for type $k$ item in round $t.$ DSP uses algorithm~$\mathcal{A}$ to schedule one of the advertisers in each round of auctions. Let $\{a_t\}_{t=1}^{T}$ be the  scheduled sequence of advertisers according to $\mathcal{A}$ and $a_t \in \mathcal{N}.$ The strategy of the DSP is DSIC. 
  The expected revenue of the DSP under algorithm~$\mathcal{A}$ over $T$ rounds of auctions is 
\begin{eqnarray}
R_{D,PPA}^{\mathcal{A}} = \mathbb{E}^{\mathcal{A}} \left[\sum_{t=1}^{T} \sum_{k=1}^{K} (v_{k,a_t} - q_{k,t}) X_{k,t} 1_{\{(q_{k,t} < v_{k,a_t}) \}} \right]. \nonumber 
\label{eqn:rppa-DSP}
\end{eqnarray}

The expected revenue of the seller over $T$ rounds of auctions when DSP uses scheduling algorithm~$\mathcal{A}$  is given by 
\begin{eqnarray}
R_{S,PPA}^{\mathcal{A}} = \mathbb{E}^{\mathcal{A}} \left[\sum_{t=1}^{T} \sum_{k=1}^{K} q_{k,t} X_{k,t} 1_{\{(q_{k,t} < v_{k,a_t}) \}} \right]. \nonumber 
\label{eqn:rppa-DSP-S}
\end{eqnarray}

The objective of DSP is to come up with a scheduling algorithm to meet the criteria of the advertisers. The goal of the seller is to set the optimal reserve price to maximize its revenue.


\section{Analysis} 

In this section, we first analyze the repeated posted price auction mechanism with a single buyer (advertiser). We derive some properties of RPPAs. Later we study  RPPAs involving a DSP.  

\subsection{Repeated posted price auction with single buyer} 
\subsubsection{Homogeneous RPPA}
 In a homogeneous RPPA the seller auctions identical items in all rounds. 
We begin with an example; suppose that the buyer has fixed valuation $v$ for each PPA (round). The best response of the buyer to the seller's reserve price $q,$ is to accept the offer if $v>q,$ and reject the offer otherwise. In each round of auctions, the buyer follows this strategy. The total revenue of the buyer over $T$ rounds is $(v-q)T 1_{\{v>q\}}.$ The seller's revenue is $qT1_{\{v>q\}}.$ In this example the buyer does not have incentive to lie and accept the offer when $v < q,$ because  the payoff is negative and  the seller uses fixed reserve price for all $T$ auctions. This strategy of the buyer is called dominant strategy incentive compatible (DSIC).  

When the buyer's valuation in each round is drawn from probability distribution $F$ over $[\underline{v}, \overline{v}].$ Each round is an independent PPA; hence, the buyer's strategy is to accept the offer when current valuation is higher than the reserve price of the seller. We make the following assumption on probability distribution function $F.$  

\begin{definition}[Assumption]
	The distribution function $F$  with density $f$ is regular and $\left(v -\frac{1-F(v)}{f(v)}\right)$ is strictly increasing in $v$ over $[\underline{v}, \overline{v}].$ 
	\label{def:regularF}
\end{definition} 
We present the following proposition which is simple variant of \cite{Myerson81}.
\begin{proposition}
In homogeneous RPPA, the seller sets a single optimal reserve price $q^*.$ The optimal revenue of  the seller over $T$ rounds of auctions is 
\begin{equation}
 T q^*\mathrm{Pr}(v > q^*),
 \label{eqn:seller-revenue-homo}
\end{equation}
and the optimal revenue of the buyer over $T$ rounds of auction is 
\begin{equation}
T \mathbb{E}[(v-q^*)1_{\{v > q^*\}}].
\label{eqn:buyer-revenue-homo}
\end{equation}

\end{proposition}
\textbf{Sketch of the proof:}
In repeated PPA, each round is PPA, where seller can determine the optimal reserve price and it has a unique solution because $F$ is regular. Taking derivative of $q(1-F(q))$ with respect to $q,$  and equating it to $0,$ we get the optimal reserve price $q^*.$ This is a solution of  
\begin{eqnarray}
  q - \frac{1 - F(q)}{f(q)} =0 .
  \label{eqn:opt-sol-reserve-price}
\end{eqnarray}
Buyer accepts PPA with probability $\mathrm{Pr}(v > q^*).$  
Then, we obtain optimal revenue of the seller from $T$ rounds of auctions as $Tq^*\mathrm{Pr}(v > q^*).$

The buyer accepts an offer if value $v_t$ drawn from $F$ in auction $t$ is higher than $q^*.$ The identical items are auctioned in round. 
 Thus, the optimal revenue of the buyer is $T \mathbb{E}[(v-q^*)1_{\{v > q^*\}}].$ 
 

\subsubsection{Heterogeneous RPPA}
 Here, the seller has different types of items for auction. 
In each round one type of item is auctioned. Both the seller and buyer know the type of item that is put in for auction. 
Our objective here is to show that, in heterogeneous RPPA, the optimal strategy for a seller is to have dynamic reserve price.  To gain insight into this, let us look at the following example. 

\textbf{Example:} Consider a buyer and seller model with two types of items to be auctioned  by the seller.  Suppose that $v_1$ and $v_2$ are valuations of the buyer for these two  types of items; and $v_1 < v_2.$ In auction $t,$ the item of type $i$ is auctioned with prob.  $p_i,$ and $0 <p_1, p_2 <1,$ $p_1 + p_2 = 1.$  Now suppose that the seller sets a reserve price $q_1,$ for all $T$ round of auctions and $q_1 < v_1 < v_2.$ 
 The revenue of the seller over $T$ rounds of auction under reserve price $q_1$ is that $q_1T.$  If seller sets reserve price higher than $v_2,$ then clearly seller's revenue is $0.$ If the seller sets the reserve price $q_2$ for all $T$ and  $v_1 < q_2 < v_2,$ then clearly there is no revenue of the seller from type $1$ items auction because reserve price is higher than valuation but the seller can have revenue from type $2$ item. Thus the  revenue of the seller is $q_2 p_2 T.$ Now suppose the seller has a  dynamic reserve price such that, when item of type $i$ is auctioned, the reserve price set is $q_i,$  and $q_1 <v_1 < q_2< v_2.$ Total revenue of the seller under dynamic reserve price is $(p_1q_1 + p_2 q_2)T.$ Clearly, we have $(p_1q_1 + p_2 q_2)T > q_2 p_2 T$ and $(p_1q_1 + p_2 q_2)T > q_1  T.$ Thus, the seller uses dynamic reserve price for multi-type of items.

We now extend this to $K$-different types of items. Their valuation is drawn from distribution $F_k$ with density $f_k$ over support $[\underline{v}_k, \overline{v}_k]$ and $\underline{v}_{k-1}<\overline{v}_k.$ 

\begin{proposition}
In heterogeneous RPPA, the optimal reserve price of the seller is dynamic. The optimal revenue of a seller using dynamic reserve price over $T$ rounds of auctions is 
\begin{eqnarray*}
	\sum_{t=1}^{T} \sum_{k=1}^{K} p_k q^{\ast}_k \mathrm{Pr}(v_{t} > q_k^{\ast} ).
\end{eqnarray*}
\end{proposition}

\textbf{Sketch of the proof:} In auction $t$ item of type $k$ is auctioned with probability $p_k,$ $\sum_{k=1}^{K}p_k =1.$ The buyer has valuation $v_t$ which is drawn from probability distribution $F_k$ over $[\underline{v}_k, \overline{v}_k]$, given that type $k$ item is auctioned. The optimal reserve price set by the seller is $q_k^*$ which is solution of $q_k - \frac{1 - F_k(q_k)}{f_k(q_k)} =0.$ The seller's revenue conditioned on type $k$ item is $q_k^*\mathrm{Pr}_{F_k}(v > q_k^*).$ For $K$ different type of items, there are different optimal reserve prices. $q^* =[q_1^*, \cdots, q^*_K] $ is vector of optimal reserve prices for $K$ type of items. The optimal revenue from a auctions is
\begin{equation*}
\sum_{k=1}^{K} p_k q_k^*\mathrm{Pr}_{F_k}(v_t > q_k^*).
\end{equation*}
  Thus total revenue over $T$ rounds of auctions is  
  \begin{eqnarray*}
  	\sum_{t=1}^{T} \sum_{k=1}^{K} p_k q^{\ast}_k \mathrm{Pr}_{F_k}(v_{t} > q_k^{\ast} ).
  \end{eqnarray*}




From the buyer's perspective, the revenue from auction $t$ conditioned on the auction of item of type $k,$ is $\mathbb{E}_{F_k}\left[(v_t - q_k^*)1_{\{v_t > q_k^*\}} \right]$
Therefore, buyer's total revenue over $T$ rounds is 
\begin{eqnarray*}
	\sum_{t=1}^{T} \sum_{k=1}^{K} p_k \mathbb{E}_{F_k}\left[(v_t - q_k^*)1_{\{v_t > q_k^*\}} \right].
\end{eqnarray*}
\subsection{Repeated posted price auction with a DSP} 
\subsubsection{Homogeneous RPPA}
There is a seller and a DSP. The seller auctions identical items in each round of auctions. DSP has $N$ advertisers (clients). 
We first study the DSP's problem of advertiser scheduling, assuming perfect hindsight of valuation for advertisers. $v_{n,t}$ is the valuation of advertiser $n$ and $q^*$ is the reserve price set by the seller. A naive scheduling scheme based on highest valuation is described in Algorithm.~\ref{algo:naive-scheme}. The advertiser with the highest valuation in round $t$ is scheduled if it is higher than reserve price $q^*.$  We denote the  allocation matrix as $\{x_{n,t}\};$   $x_{n,t} = 0$ if $v_{n,t} < q^*,$ for $ n \in \mathcal{N},$ $1 \leq t \leq T$ and ${x}_{n,t} \in \{0,1\}$  if $v_{n,t} \geq q^*,$ for $n\in \mathcal{N},$ $1 \leq t \leq T.$ If two or more advertisers share the highest valuation, then one of them is scheduled randomly.

\begin{algorithm}[h]
	\KwIn{The valuations of the advertisers, $v_{n,t}$ for all $n \in \mathcal{N},$ $1 \leq t \leq T$ and seller's reserve price $q^*.$  }
	\KwOut{Allocation matrix $x_{n,t}$}
	\For {$1 \leq t \leq T$}{
		$v_{\max,t} =  \max_{n \in \mathcal{N}} v_{n,t}$ \\
		$a_t = \arg \max_{n \in \mathcal{N}} v_{n,t}$ \\
		$x_{n,t} = 1$  \mbox{if $a_t = n,$ $\&$ $v_{\max,t} \geq  q^*$ }\\
		 $x_{n,t} = 0$ \mbox{else} \\
		\textbf{end}
		
	} 
	
	\caption{{\sc DSP} Scheduling algorithm for advertisers. }
	\label{algo:naive-scheme}
\end{algorithm}

When the valuation of the advertiser $v_{n,t} = v$ for all $1 \leq t \leq T$ and all $n \in \mathcal{N}$ then advertisers are scheduled uniform randomly. If $v_{1,t} > \max_{n \in \mathcal{N}, n \neq 1} v_{n,t},$ for all $1 \leq t \leq T,$ then according to Algorithm~\ref{algo:naive-scheme}, DSP schedules only advertiser $1.$  Algorithm~\ref{algo:naive-scheme} is not a suitable choice for the DSP as it is not fair; some advertisers are disproportionately favored. Advertisers often have impression constraints, i.e., they want to display their ad for a fixed number of times. To account for this, we formulate the following problem under perfect hindsight of valuation of advertiser. 
\begin{equation}
\label{eq:opt-prob}
\tag{P1} 
\begin{array}{rll}
\max_x  & & G(x) = \sum_{t=1}^{T} \sum_{n=1}^{N} \left[ (v_{n,t} - q^*)x_{n,t}\right] \\
\mbox{s.t.} & & \sum_{n = 1}^{N} {x}_{n,t} \leq  1, \text{ for } 1 \leq t \leq T\\ 
& &  \sum_{t=1}^{T}  \sum_{n = 1}^{N} {x}_{n,t} \leq  T \\
& &  \sum_{t=1}^{T}   {x}_{n,t} \geq  \Delta_n \ \ \mbox{ for $n\in \mathcal{N}$ } \\ 
& & {x}_{n,t} = 0 \ \ \mbox{if $v_{n,t} < q^*,$ for $ n \in \mathcal{N},$ $1 \leq t \leq T$} \\
& & {x}_{n,t} \in \{0,1\}  \ \ \mbox{if $v_{n,t} \geq q^*,$ for $n\in \mathcal{N},$ $1 \leq t \leq T.$} \\
\end{array}
\end{equation}

Note that problem \eqref{eq:opt-prob} is a linear program with integer constraints. The first constraint is that only an advertiser can be scheduled, the second is a  constraint on the total number of auctions won by the DSP, the third constraint  presents the demand of impressions from the advertiser, $\Delta_n,$  we assume that $\Delta_n < T.$ Fourth and fifth constraints are on the values of allocation variable $x_{n,t}.$ The solution of the problem can be approached using Lagrangian relaxation method. This provides an upper bound on the solution of problem \eqref{eq:opt-prob}, \cite[Chapter $10$]{Bertsekas98}. One can further relax the integer constraint; this is the relaxed Lagrangian problem. Then, a subgradient scheme can be used to update the Lagrangian variable. 

We now  suppose that the valuations of advertisers are drawn from identical distribution $F$ with density $f$ and follow the regularity property. Then, the seller can set optimal reserve price $q^*.$ Note that this reserve price is fixed for all the rounds of auctions.  The DSP sequentially schedules an advertiser from the set of advertisers. It solves the following optimization problem. 
\begin{equation}
\label{eq:opt-prob2}
\tag{P2} 
\begin{array}{rll}
\max_x  & &  \sum_{t=1}^{T} \sum_{n=1}^{N} \mathbb{E}_{F}  \left[ (v_{n,t} - q^*) 1_{\{v_{n,t} > q^*\} } x_{n,t}\right]  \\
\mbox{s.t.} & & \sum_{n = 1}^{N} {x}_{n,t} \leq  1, \text{ for } 1 \leq t \leq T\\ 
& & {x}_{n,t} = 0 \ \ \mbox{ with prob. $F(q^*)$ for all  $n, t$} \\
& & {x}_{n,t} \in \{0,1\}  \ \ \mbox{with prob. $1-F(q^*)$ for all  $n, t$} \\
\end{array}
\end{equation}
In this optimization formulation, the expectation is w.r.t. distribution $F.$  The  first constraint indicates that only one advertiser can be scheduled in each round. The condition in the second constraint implies that $v_{n,t} < q^*$ and hence $x_{n,t} =0.$ This occurs with prob. $F(q^*).$  The third condition implies that $v_{n,t} \geq q^*$ and hence $x_{n,t} \in \{0,1\}.$ This occurs with prob. $1-F(q^*).$  

A simple strategy for the DSP would be to schedule the advertiser with maximum valuation in the current round if it is higher than optimal reserve price, otherwise reject the offer of the seller.  This allocation mechanism is randomized because  the valuation of each advertiser is drawn according to $F$ from $[\underline{v}, \overline{v}].$ This is described in Algorithm~\ref{algo:naive-scheme-random}. 

\begin{algorithm}[h]
	\KwIn{Seller's reserve price is $q^*.$  }
	\KwOut{Allocation matrix $x_{n,t}$}
	\For {$1 \leq t \leq T$}{
		Draw $v_{n,t}$ according to $F$ from $[\underline{v}, \overline{v}]$ \\
		$v_{\max,t} =  \max_{n \in \mathcal{N}} v_{n,t}$ \\
		$a_t = \arg \max_{n \in \mathcal{N}} v_{n,t}$ \\
		$x_{n,t} =1$	if	$a_t =n$ and $ v_{max,t} > q^*$ \\
		$x_{n,t} =0$ else
	} 
	
	\caption{{\sc DSP}'s randomized scheduling algorithm for advertisers. }
	\label{algo:naive-scheme-random}
\end{algorithm} 
The probability of scheduling an advertiser in  round $t$ is given by 
{\small{
\begin{eqnarray*}
\mathrm{Pr}( \max_n v_{n,t}  > q^*) &=&  1 - \mathrm{Pr}( \max_n v_{n,t}  \leq q^*) \\
&=& 1- \prod_{n=1}^{N} \mathrm{Pr}( v_{n,t}  \leq q^*) \\
&=& 1- (F(q^*))^N.
\end{eqnarray*} }}
The expected number of impressions (auctions) won over $T$ rounds of auctions is $T (1- (F(q^*))^N).$ Hence, the expected number of impressions assigned to an advertiser is $\frac{T (1- (F(q^*))^N)}{N}.$ 

The advertisers may also have different demands for impressions. Thus, the DSP may need to consider the following impression constraint optimization problem, where constraints are on expected impressions. 
{\small{
\begin{equation}
\label{eq:opt-prob3}
\tag{P3} 
\begin{array}{rll}
\max_x  & &    \sum_{t=1}^{T} \sum_{n=1}^{N} \mathbb{E}_{F} \left[ (v_{n,t} - q^*) 1_{\{v_{n,t} > q^*\} } x_{n,t}\right]  \\
\mbox{s.t.} & & \sum_{n = 1}^{N} {x}_{n,t} \leq  1, \text{ for } 1 \leq t \leq T\\ 
& &  \mathbb{E}_F\left[\sum_{t=1}^{T}  1_{\{v_{n,t} > q^*\}}  {x}_{n,t} \right] \geq  \Delta_n \ \ \mbox{ for $n\in \mathcal{N}$ } \\ 
& & {x}_{n,t} = 0 \ \ \mbox{ with prob. $F(q^*)$ for all  $n, t$} \\
& & {x}_{n,t} \in \{0,1\}  \ \ \mbox{with prob. $1-F(q^*)$ for all  $n, t.$} \\
\end{array}
\end{equation} }}
Here, second condition gives the expected demand constraints from advertisers. Again, solution to problem \eqref{eq:opt-prob3} can be obtained using Lagrangian relaxation approach. 
 
\subsubsection{Heterogeneous RPPA}
We now analyze  the DSP's advertiser scheduling problem for a heterogeneous RPPA. At the beginning of each round, the DSP and seller know the type of item  that is put in for auction. Using this information, the seller sets an  optimal reserve price and it is dependent on type of item. 

We assume that DSP has  perfect hindsight of valuation for advertisers. Let $v_{n,k,t}$ be the valuation of advertiser $n$ if item of type $k$ is auctioned in round $t.$ Let $q^*_k$ be the optimal reserve price for type $k$ item, set by the  seller. As before, we first study a naive scheduling scheme based on highest valuation; this is given in Algorithm.~\ref{algo:naive-scheme-hetero}.  The advertiser with the highest valuation in round $t$ is scheduled if its valuation is higher than reserve price $q^*_k.$ If two or more advertisers share the highest valuation, then one of them is scheduled randomly. We use $\{b_t\}_{t=1}^T$ to denote the sequence of the type of item auctioned; $b_t \in \{q_1^*,\cdots q_K^*\}.$

\begin{algorithm}[h]
	\KwIn{The valuations of the advertisers, $v_{n,k,t}$ for all $n \in \mathcal{N},$ $1 \leq t \leq T.$  and $\{b_t\}_{t=1}^T$  	   }
	\KwOut{Allocation matrix $x_{n,t}$}
	\For {$1 \leq t \leq T$}{
		Type of item for auction is obsereved, i.e., $k$ \\
		$v_{\max,t} =  \max_{n \in \mathcal{N}} v_{n,k,t}$ \\
		$a_t = \arg \max_{n \in \mathcal{N}} v_{n,t}$ \\
		Seller sets the optimal reserve price $b_t = q^*_k$ \\
		$x_{n,t} = 1$  \mbox{if $a_t = n,$ $\&$ $v_{\max,t} \geq  b_t$ }\\
		$x_{n,t} = 0$ \mbox{else} \\
		\textbf{end}
		
	} 
	
	\caption{{\sc DSP} Scheduling algorithm of heterogeneous items for advertisers. }
	\label{algo:naive-scheme-hetero}
\end{algorithm}

The advertisers can have constraints on number of impressions. Also, each advertiser may prefer different types of items with different frequency. Hence, the  DSP considers the following optimization formulation. 

{\footnotesize{
\begin{equation}
\label{eq:opt-prob4}
\tag{P4} 
\begin{array}{rll}
\max_x  & &  \sum_{t=1}^{T} \sum_{n=1}^{N} \sum_{k=1}^{K} \left[ (v_{n,k,t} - b_t)x_{n,t}\right] \\
\mbox{s.t.} & & \sum_{n = 1}^{N} {x}_{n,t} \leq  1, \text{ for } 1 \leq t \leq T\\ 
& & \sum_{t=1}^{T} 1_{\{v_{n,k,t} > q^*_k \}}x_{n,t} \geq y_{n,k}  \mbox{ for $n \in \mathcal{N},$ $1 \leq t \leq T$ }\\
& &  \sum_{t=1}^{T}   {x}_{n,t} \geq  \Delta_n \ \ \mbox{ for $n\in \mathcal{N}$ } \\ 
& & {x}_{n,t} = 0 \ \ \mbox{if $v_{n,k,t} < b_t,$ for $ n \in \mathcal{N},$ $1 \leq t \leq T$} \\
& & {x}_{n,t} \in \{0,1\}  \ \ \mbox{if $v_{n,k,t} \geq b_t,$ for $n\in \mathcal{N},$ $1 \leq t \leq T.$} \\
\end{array}
\end{equation}
}}

Note that all constraints in problem~\eqref{eq:opt-prob4} are similar to earlier problem~\eqref{eq:opt-prob}, except the second constraint which indicates demands of advertisers for the number of impressions for different types of items.

We next study heterogeneous RPPA when valuation of the advertisers is a realization drawn from fixed distributions. An item of type $k$ is auctioned in slot $t$ with  probability $p_k.$ The seller and DSP has knowledge of type of items at the beginning of every auction. The DSP can schedule advertisers based on the highest valuation. The probability of scheduling an advertiser in round $t$ given that type $k$ item is auctioned, is given by 
\begin{eqnarray*}
	\mathrm{Pr}( \max_n v_{n,k,t}  > q_k^*~|~ \text{ type $k$ item }) &=& 1- (F_k(q^*_k))^N.
\end{eqnarray*}
The expected number of impressions (auctions) won over $T$ rounds of auctions are $T \sum_{k=1}^{K} p_k (1- (F_k(q^*_k))^N).$ Hence, the expected number of impressions assigned to an advertiser are $\frac{T \sum_{k=1}^{K} p_k (1- (F_k(q^*_k))^N)}{N}.$ 

The advertisers may have an expectation constraint on the number on impressions. In this case the DSP considers the following impression constraint optimization problem. 

{\footnotesize{
\begin{equation}
\label{eq:opt-prob5}
\tag{P5} 
\begin{array}{rll}
\max_x   &   \left[ \sum_{t=1}^{T} \sum_{n=1}^{N} \sum_{k=1}^{K} p_k \mathbb{E}_{F_k} \left[ (v_{n,k,t} - q^*_k) 1_{\{v_{n,k,t} > q_k^*\} } x_{n,t}\right] \right]  \\
\mbox{s.t.}  & \sum_{n = 1}^{N} {x}_{n,t} \leq  1, \text{ for } 1 \leq t \leq T\\ 
& \mathbb{E}_{F_k}\left[\sum_{t=1}^{T} 1_{\{v_{n,k,t} > q^*_k \}}x_{n,t} \right]= y_{n,k}  \mbox{ for $n \in \mathcal{N},$ $1 \leq t \leq T$ } \\
 &  \sum_{t=1}^{T} \sum_{k=1}^{K} \mathbb{E}_{F_k}\left[  1_{\{v_{n,k,t} > q^*_k\}}  {x}_{n,t} \right] \geq  \Delta_n \ \ \mbox{ for $n\in \mathcal{N}$ } \\ 
 & {x}_{n,t} = 0 \ \ \mbox{ with prob. $\sum_{k=1}^{K}p_k F_k(q^*_k)$ for all  $n, t$} \\
 & {x}_{n,t} \in \{0,1\}  \ \ \mbox{with prob. $1- \sum_{k=1}^{K}p_k F_k(q^*_k))$ for all  $n, t.$} \\
\end{array}
\end{equation}
}}
Here, conditioned on type of item $k$ is auctioned in round $t$, $x_{n,t} =0$ with prob. $F_k(q_k^*).$ Hence unconditional this, $x_{n,t} = 0$ with prob. $ \sum_{k=1}^{K}p_k F_k(q^*_k).$  Similarly, $x_{n,t} \in \{0,1\}$ with prob. $1- \sum_{k=1}^{K}p_k F_k(q^*_k).$ 


\section{Budget constrained advertisers}

Let us now look at the budget management strategy for advertiser, i.e., buyer's perspective in posted price auction.  Here, the seller sets a fixed reserve price, say  $q$.
To gain insight into this, consider a single advertiser and seller model with posted price mechanism. Let $v$ be the fixed valuation of advertiser with total budget $B$ and he is interacting with the seller over $T$ rounds. Let  $v > q.$ 
 If total budget $B > Tq,$  then the advertiser can win all impressions. Suppose that $B < Tq,$ the advertiser can win a fraction $M$ of impressions from $T,$ i.e., $M =\frac{B}{q}.$ A simple strategy of the advertiser could be to buy the first $M$ impressions out of $T.$ But, this may not be suitable for an advertiser who wants  to display ads uniformly over $T$ auctions. Hence, the advertiser uses a  probabilistic throttling scheme in which he participates in each round of auctions   with probability $\xi,$ and $0<\xi < \frac{M}{T}<1.$ 
 This simple scenario suggests that an advertiser with tight budget would prefer to participate in auctions selectively.
 
 Now consider a DSP with clients who are budget constrained.
Let us now look at the problem of budget-constrained-advertiser scheduling at the DSP in a homogeneous RPPA. Also, the advertisers have distinct budget constraints.  Let $B_n$ be the budget of advertiser $n.$ For simplicity assume that all advertisers have the same valuation $v$ and this is higher than seller's reserve price $q.$ And, $\sum_{n = 1}^{N} B_n \leq Tq.$ The DSP can use probabilistic throttling scheme in which it schedules advertiser $n$ in each round with probability $0<\xi_n < \frac{B_i}{q}<1,$ and  $\sum_{n = 1}^N\xi_n =1.$

For the case where valuation changes for every round, the DSP considers following  budget constrained optimization problem under perfect hindsight of valuation of advertisers.
 \begin{equation}
 \label{eq:opt-prob6}
 \tag{P6} 
 \begin{array}{rll}
 \max_x  & & \sum_{t=1}^{T} \sum_{n=1}^{N} \left[ (v_{n,t} - q^*)x_{n,t}\right] \\
 \mbox{s.t.} & & \sum_{n = 1}^{N} {x}_{n,t} \leq  1, \text{ for } 1 \leq t \leq T\\ 
 & &  \sum_{t=1}^{T}   {x}_{n,t} \leq  \frac{B_n}{q^*} \ \ \mbox{ for $n\in \mathcal{N}$ } \\ 
 & & {x}_{n,t} = 0 \ \ \mbox{if $v_{n,t} < q^*,$ for $ n \in \mathcal{N},$ $1 \leq t \leq T$} \\
 & & {x}_{n,t} \in \{0,1\}  \ \ \mbox{if $v_{n,t} \geq q^*,$ for $n\in \mathcal{N},$ $1 \leq t \leq T.$} \\
 \end{array}
 \end{equation}

Note that in the above formulation, $B_n$ is budget of advertiser $n,$ and the second constraint indicates the budget constraint of that advertiser. 

Next, suppose that the valuations of advertisers are drawn from identical distribution $F$ with density $f$ and seller sets the optimal reserve price $q^*.$  The DSP sequentially schedules budget constrained advertisers by solving the following optimization problem. 
\begin{equation}
\label{eq:opt-prob7}
\tag{P7} 
\begin{array}{rll}
\max_x  & &  \mathbb{E}_{F} \left[ \sum_{t=1}^{T} \sum_{n=1}^{N} \left[ (v_{n,t} - q^*) 1_{\{v_{n,t} > q^*\} } x_{n,t}\right] \right]  \\
\mbox{s.t.} & & \sum_{n = 1}^{N} {x}_{n,t} \leq  1, \text{ for } 1 \leq t \leq T\\ 
& & \mathbb{E}_F\left[\sum_{t=1}^{T} 1_{\{v_{n,t} > q^*\}}   {x}_{n,t} \right]  \leq  \frac{B_n}{q^*} \ \ \mbox{ for $n\in \mathcal{N}$ } \\
& & {x}_{n,t} = 0 \ \ \mbox{ with prob. $F(q^*)$ for all  $n, t$} \\
& & {x}_{n,t} \in \{0,1\}  \ \ \mbox{with prob. $1-F(q^*)$ for all  $n, t$} \\
\end{array}
\end{equation}
 Second  constraint indicates that the expected budget constraint on each advertiser.

In heterogeneous RPPA, single seller and an advertiser problem is non trivial. The advertiser may have preferences for different type of items. Hence he would like balance budget for distinct items. This is again constrained optimization problem. 
Finally, we consider DSP with budget constrained advertisers and DSP  solves the following budget constrained optimization problem.

{\footnotesize{
		\begin{equation}
		\label{eq:opt-prob8}
		\tag{P8} 
		\begin{array}{rll}
		\max_x   &   \sum_{t=1}^{T} \sum_{k=1}^{K} p_k \sum_{n=1}^{N} \mathbb{E}_{F_k} \left[ (v_{n,k,t} - q^*_k) 1_{\{v_{n,k,t} > q_k^*\} } x_{n,t}\right]  \\
		\mbox{s.t.}  & \sum_{n = 1}^{N} {x}_{n,t} \leq  1, \text{ for } 1 \leq t \leq T\\ 
		& \mathbb{E}_{F_k}\left[\sum_{t=1}^{T} 1_{\{v_{n,k,t} > q^*_k \}}x_{n,t} \right] \geq y_{n,k}  \mbox{ for $n \in \mathcal{N},$ $1 \leq t \leq T$ } \\
		&  \sum_{t=1}^{T} \sum_{k=1}^{K} p_k\mathbb{E}_{F_k}\left[ q_k^*  1_{\{v_{n,k,t} > q^*_k\}}  {x}_{n,t} \right] \leq  B_n \ \ \mbox{ for $n\in \mathcal{N}$ } \\ 
		& {x}_{n,t} = 0 \ \ \mbox{ with prob. $\sum_{k=1}^{K}p_k F(q^*_k)$ for all  $n, t$} \\
		& {x}_{n,t} \in \{0,1\}  \ \ \mbox{with prob. $\sum_{k=1}^{K}p_k (1-F(q^*_k))$ for all  $n, t.$} \\
		\end{array}
		\end{equation}
}}

We note that in above formulation, second constraint indicated different demands for different type of items and the third constraint represent the expected budget constraint.

\section{Numerical results}
\label{subsec:Num-result-TS}
%

In this section, we first provide numerical examples for a single seller and single buyer model. We obtain the optimal reserve price when the valuation of the buyer is a random variable with uniform , exponential, and lognormal distribution. Also, we  compute the expected revenue of the buyers over $T$ rounds of auctions. Next we consider a numerical example for the DSP  under impression constraints.  Finally, we provide simulations for a DSP with multiple budget constraint advertisers. 

\begin{itemize}
	\item Uniform distribution $U[\underline{v},\overline{v}],$ then cdf is
	\begin{eqnarray*}
	F(v) = \begin{cases}
	0 \ \mbox{if $v < \underline{v}$} \\
	\frac{v- \underline{v}}{\overline{v} - \underline{v}} \ \mbox{if $v \in [\underline{v}, \overline{v})$} \\
	1 \ \mbox{if $v > \overline{v}.$}
	\end{cases}
	\end{eqnarray*}
	and pdf is 
	\begin{eqnarray*}
	f(v) = \begin{cases}
	\frac{1}{\overline{v} - \underline{v}} \ \mbox{if $v \in [\underline{v}, \overline{v}]$} \\
	0 \ \mbox{Otherwise}
	\end{cases}
	\end{eqnarray*}
	Then using eqn.~\eqref{eqn:opt-sol-reserve-price}, we obtain 
	\begin{eqnarray*}
	v- \frac{1- \frac{v- \underline{v}}{\overline{v} - \underline{v}} }{\frac{1}{\overline{v} - \underline{v}}} = 0.
	\end{eqnarray*} 
	and hence we can have optimal reserve price $q^* = \frac{\overline{v}}{2}.$ In the example $U[0,1],$ optimal reserve price $1/2.$ From Eqn.~\eqref{eqn:seller-revenue-homo} and \eqref{eqn:buyer-revenue-homo}, the expected revenue of a seller is $\frac{T}{4}$ and expected revenue of a buyer is $\frac{T}{8}.$ 
	\item Exponential distribution $Exp[\lambda],$ then cdf is
	\begin{eqnarray*}
	F(v) = 1- e^{-\lambda v} \ \mbox{for $v \in [0, \infty)$}.
	\end{eqnarray*}
	and pdf is 
	\begin{eqnarray*}
	f(v) = \lambda e^{-\lambda v}
	\end{eqnarray*}
	Then using eqn.~\eqref{eqn:opt-sol-reserve-price}, we obtain 
	\begin{eqnarray*}
	v- \frac{1-( 1- e^{-\lambda v})}{\lambda e^{-\lambda v}} = 0.
	\end{eqnarray*} 
	and hence we can have optimal reserve price $v^* = \frac{1}{\lambda}.$ 
	Then 
{\small{
	\begin{eqnarray*}
	\mathrm{Pr}(v^*>\frac{1}{\lambda} ) = 1- \mathrm{Pr}(v^*\leq \frac{1}{\lambda} ) \\
	= 1- \left( 1 - e^{-\lambda \times \frac{1}{\lambda}}\right) 
	= 1 - 1 + e^{-1} 
	= e^{-1}
	\end{eqnarray*}
	}}
	Thus the seller's revenue is $T/ \lambda e.$ Buyers expected revenue is 
{\small{
	\begin{eqnarray*}
	T	\int_{1/\lambda}^{\infty} \left( x - \frac{1}{\lambda}\right) \lambda e^{-\lambda x} dx = \frac{T}{\lambda e}.
	\end{eqnarray*} }}
	Note that mean of exponential random variable is $\mu = \frac{1}{\lambda}.$ This suggest that the higher the mean implies higher expected revenue for both buyer and seller.
	
	\item Lognormal distribution: The cdf with mean $\mu$ and variance $\sigma$ is 
	\begin{eqnarray*}
	F(v) = \Phi \left( \frac{\ln v - \mu}{\sigma} \right),
	\end{eqnarray*}	 
	where  $\Phi$ is the cumulative distribution function of the standard normal distribution (i.e. $N(0,1)$). 
	The probability density function is 
	\begin{eqnarray*}
	f(v) = \frac{1}{v \sigma \sqrt{2 \pi}} \exp{\left(-\frac{(\ln v - \mu)^2}{2 \sigma^2}\right)}.
	\end{eqnarray*}
	Then using eqn.~\eqref{eqn:opt-sol-reserve-price}, we get {\small{
	\begin{eqnarray*}
	v- \frac{1- \Phi \left( \frac{\ln v - \mu}{\sigma} \right)}{\frac{1}{v \sigma \sqrt{2 \pi}} \exp{\left(-\frac{(\ln v - \mu)^2}{2 \sigma^2}\right)}} = 0.
	\end{eqnarray*} }}
	After further simplification  we get 
{\small{	\begin{eqnarray*}
		\frac{1}{ \sigma \sqrt{2 \pi}} \exp{\left(-\frac{(\ln v - \mu)^2}{2 \sigma^2}\right)} - 1 + \Phi \left( \frac{\ln v - \mu}{\sigma} \right) = 0.
	\end{eqnarray*} }}
	Analytical solution of this equation is difficult to obtain. We compute the  solution using numerical methods and get the optimal reserve price $q^*.$  Buyers expected revenue is {\small{
	\begin{eqnarray*}
		T \int_{q^*}^{\infty} (v-q^*) \frac{1}{\sigma v \sqrt{2\pi}} \exp\left( -\frac{(\ln v - \mu )^2}{2 \sigma^2}  \right) dv.
	\end{eqnarray*} }}
	Seller's expected revenue is 
{\small{	\begin{eqnarray*}
		T q^* \left( 1 - \Phi \left( \frac{\ln v - \mu}{\sigma} \right)\right). 
	\end{eqnarray*} }}
	It is difficult to write the above expressions in closed form. Hence, we provide  results through numerical methods, in Table~\ref{table:Buyer-seller-model}.  
	\begin{table}[h]
		\centering
		\caption{The optimal reserve price, expected  revenue of seller (Seller R) and expected revenue of buyer (Buyer R) for different $\mu.$  }
		\label{table:Buyer-seller-model}
		\begin{tabular}{|c|c|c|c|c|}
			\hline
			$\mu$   & $\sigma$      & $q^*$    & Seller R  & Buyer R \\ \hline
			$0$  & $1$  & $1.4$ & $0.5156T$  & $0.7512T$\\ \hline
			$0.25$  & $1$  & $1.8$ & $0.8380T$  & $0.9175T$\\ \hline
			$0.5$  & $1$  & $2.3$ & $1.23T$  & $1.18 T$\\ \hline	
			$2$  & $1$  & $10.1$ & $9.61T$  & $5.37T$\\ \hline
		\end{tabular}
	\end{table}
	From Table~\ref{table:Buyer-seller-model} observe that as mean $\mu$ increases, the optimal reserve price of the seller increases. This means, the sellers expected revenue also increases. Further, the expected revenue of buyer also increases with  $\mu.$   
\end{itemize}


{ 
\subsection{DSP Problem}
\label{sec:numerical-DSP-problem}
We present numerical example for RPPA with a DSP, and use the naive scheduling scheme (see Algorithm~\ref{algo:naive-scheme})  when the valuations of advertisers are known in hindsight. 
We use following parameters: number of auctions $T= 10000,$ $N =5.$ We use fixed valuation $v_{n,t}$ and generate $[[v_{n,t}]]$ using lognormal distribution with mean $\mu=0$ and variance $\sigma=1$. For different reserve prices $q$, we provide simulations to illustrate effect of reserve price on the expected  revenue of seller ($\frac{\sum_{t=1}^{T} q1_{\{\max_{n} v_{n,t} > q\}}}{T}$), the expected revenue of buyers ($\frac{\sum_{t=1}^{T} (v_{n,t} -q)x_{n,t}}{T}$),and number of impressions win $(\sum_{t=1}^{T}x_{n,t}).$ The results are summarized in the tables below
\begin{table}[h]
\begin{center}
\begin{tabular}{|l|c|c|c|}
	\hline
 Reserve price	$q$	 & $q=1$ & $q=2$ & $q=4$ \\ \hline
 Seller's revenue& $0.9727$ & $1.5104$ & $1.3812$ \\ \hline
\end{tabular}
\end{center}
\end{table}
\begin{table}[h]
\begin{center}
\begin{tabular}{l|c|c|c|c|c|c|}
\cline{2-7}
                           & \multicolumn{2}{c|}{$q=1$} & \multicolumn{2}{c|}{q=2} & \multicolumn{2}{c|}{q=4} \\ \hline
\multicolumn{1}{|l|}{Adv.} & Impressions         & Revenue        & Imp.        & Rev        & Imp.        & Rev.       \\ \hline
\multicolumn{1}{|l|}{1}    & 2004         &  $0.6449$   & 1551        & $0.4651$           & 707         &  $0.2506$  \\ \hline
\multicolumn{1}{|l|}{2}    & 1906         &  $0.5951$   & 1484        & $0.4226$           & 683         &  $0.2137$  \\ \hline
\multicolumn{1}{|l|}{3}    & 1882         &  $0.5759$   & 1457        & $0.4060$          & 643         &  $0.2050$  \\ \hline
\multicolumn{1}{|l|}{4}    & 1950         &  $0.6125$   & 1517        & $0.4363$   & 735         &  $0.2221$  \\ \hline
\multicolumn{1}{|l|}{5}    & 1985         &  $0.6239$   & 1543        & $0.4454$           & 685         &  $0.2316$  \\ \hline
\end{tabular}
\end{center}
\end{table}
%
%
The above numerical examples suggests that, setting of optimal reserve price by the  seller impacts its expected revenue. High reserve price may lead to low revenue. And, too low reserve price also leads to low revenue. But by setting optimal reserve price, a seller can optimize its revenue.  As reserve price of seller increases, the number of impressions won by DSP for advertiser decreases and also the expected revenue of advertisers decreases. 
}

More simulation results  are provided in \cite{Meshram19}.

\section{Concluding Remarks}
This paper has studied repeated posted price auction (RPPA) in online advertising. We considered the problem of a single seller and an advertiser where items are auctioned sequentially. This was studied for both homogeneous and heterogeneous items. We also computed the optimal reserve price for seller under homogeneous items. We have shown that for heterogeneous items, the seller's optimal reserve price is dynamic. Later, we studied an RPPA involving a DSP, and devised a simple  scheduling scheme under assumption of hindsight on valuation for repeated auctions. We also formulated the advertiser scheduling problem of the DSP when the advertisers have  impression and budget constraints. This was again done for both homogeneous and heterogeneous items.  Finally, we presented some numerical examples for a single seller and advertiser model with PPA and considered various valuation distributions such as uniform, exponential and lognormal. We also provided numerical examples for a single seller and DSP model with  a naive scheduling scheme based on valuation of advertisers.

\bibliographystyle{IEEEbib}

\bibliography{ad-auction}

\section{Appendix}

\subsection{Numerical Example and Extension of Table~\ref{table:Buyer-seller-model}}
In this, we provide a numerical example of a single seller and advertiser model with lognormal distribution function for different values of variance $\sigma.$ This is given in Table~\ref{table:Buyer-seller-model-2}. 

\begin{table}[h]
	\centering
	\caption{The optimal reserve price, expected  revenue of seller (Seller R) and expected revenue of buyer (Buyer R) for different $\sigma.$  }
	\label{table:Buyer-seller-model-2}
	\begin{tabular}{|c|c|c|c|c|}
		\hline
		$\mu$   & $\sigma$      & $q^*$    & Seller R  & Buyer R \\ \hline
		$0$  & $0.25$  & $0.76$ & $0.07T$  & $0.28T$\\ \hline
		$0$  & $0.5$  & $0.78$ & $0.2T$  & $0.41T$\\ \hline
		$0$  & $1$  & $1.36$ & $0.5T$  & $0.73 T$\\ \hline	
		$0$  & $2$  & $23.2$ & $10.05T$  & $3.57T$\\ \hline
	\end{tabular}
\end{table}
We observe that as variance $\sigma$ increases,  expected revenue of the seller and advertiser increases. Further, the seller's expected revenue increase is significantly higher because the reserve price $q^*$ increase is steep. 

\subsection{Numerical Example of DSP Problem}
We here present additional numerical examples with a single seller and DSP problem. We use the same parameters as in Section~\ref{sec:numerical-DSP-problem}. Here, we study a round-robin and randomized policy for scheduling of advertiser.  

In round-robin (RR) policy, DSP selects an advertisers are selected in round-robin fashion, and are scheduled for participation in auction if their valuation is higher than the reserve price. Otherwise they are not scheduled. Numerical results are presented in Table~\ref{table:DSP-val-model-sell-rev-RR} and \ref{table:DSP-val-model-adv-rev-RR}. Since, in some auctions no advertiser is scheduled, advertisers win fewer number of impressions as compared to given in Sec.~\ref{sec:numerical-DSP-problem}. Here, as the reserve price increases, the expected revenue of the seller and of the advertisers decreases.  

\begin{table}[h]
	\begin{center}
		\caption{The seller's revenue for different reserve price when DSP uses round robin policy }
		\label{table:DSP-val-model-sell-rev-RR}
		\begin{tabular}{|l|c|c|c|}
			\hline
			Reserve price	$q$	 & $q=1$ & $q=2$ & $q=4$ \\ \hline
			Seller's revenue& $0.502$ & $0.4954$ & $0.3324$ \\ \hline
		\end{tabular}
	\end{center}
\end{table}
\begin{table}[h]
	\begin{center}
		\caption{Advertiser's revenue for different reserve price when DSP uses round robin policy}
		\label{table:DSP-val-model-adv-rev-RR}
		\begin{tabular}{l|c|c|c|c|c|c|}
			\cline{2-7}
			& \multicolumn{2}{c|}{$q=1$} & \multicolumn{2}{c|}{q=2} & \multicolumn{2}{c|}{q=4} \\ \hline
			\multicolumn{1}{|l|}{Adv.} & Impressions         & Revenue        & Imp.        & Rev.        & Imp.        & Rev.       \\ \hline
			\multicolumn{1}{|l|}{1}    &993        &  $0.1713$   & 499        & $0.1011$           & 164         &  $0.0440$  \\ \hline
			\multicolumn{1}{|l|}{2}    & 1016         &  $0.1813$   & 483       & $0.1104$           & 165         &  $0.0524$  \\ \hline
			\multicolumn{1}{|l|}{3}    & 994         &  $0.1748$   & 482       & $0.1062$          & 154         &  $0.0504$  \\ \hline
			\multicolumn{1}{|l|}{4}    & 1003         &  $0.1787$   & 514        & $0.1073$   & 175         &  $0.0481$  \\ \hline
			\multicolumn{1}{|l|}{5}    & 1014         &  $0.1825$   & 499        & $0.1115$           & 173         &  $0.0513$  \\ \hline
		\end{tabular}
	\end{center}
\end{table}
%

We now study a randomized policy. In this policy, DSP selects  advertisers in  uniform random fashion. These randomly selected advertisers are scheduled for  participation in auctions if their valuation is higher than the seller's reserve price $q.$ Numerical examples with different reserve price values are given in Table~\ref{table:DSP-val-model-sell-rev-rand} and \ref{table:DSP-val-model-adv-rev-rand}. In this policy, if DSP selects an advertiser whose valuation is smaller than the reserve price, no advertiser is scheduled in auction.  Observe that as reserve price increases, expected revenues of the seller and the advertisers decrease. 

\begin{table}[h]
	\begin{center}
		\caption{The seller's revenue for different reserve price when DSP uses uniform randomized policy }
		\label{table:DSP-val-model-sell-rev-rand}
		\begin{tabular}{|l|c|c|c|}
			\hline
			Reserve price	$q$	 & $q=1$ & $q=2$ & $q=4$ \\ \hline
			Seller's revenue& $0.493$ & $0.482$ & $0.3364$ \\ \hline
		\end{tabular}
	\end{center}
\end{table}
\begin{table}[h]
	\begin{center}
		\caption{Advertiser's revenue for different reserve price when DSP uses uniform randomized policy}
		\label{table:DSP-val-model-adv-rev-rand}
		\begin{tabular}{l|c|c|c|c|c|c|}
			\cline{2-7}
			& \multicolumn{2}{c|}{$q=1$} & \multicolumn{2}{c|}{q=2} & \multicolumn{2}{c|}{q=4} \\ \hline
			\multicolumn{1}{|l|}{Adv.} & Impressions         & Revenue        & Imp.        & Rev        & Imp.        & Rev.       \\ \hline
			\multicolumn{1}{|l|}{1}    &961        &  $0.183$   & 489        & $0.1027$           & 160         &  $0.0528$  \\ \hline
			\multicolumn{1}{|l|}{2}    & 1017         &  $0.18$   & 459        & $0.0846$           & 189         &  $0.0543$  \\ \hline
			\multicolumn{1}{|l|}{3}    & 960         &  $0.17$   & 476       & $0.1072$          & 134         &  $0.0432$  \\ \hline
			\multicolumn{1}{|l|}{4}    & 948         &  $0.16$   & 480        & $0.1033$   & 192         &  $0.0577$  \\ \hline
			\multicolumn{1}{|l|}{5}    & 1042         &  $0.185$   & 506        & $0.1054$           & 166         &  $0.0442$  \\ \hline
		\end{tabular}
	\end{center}
\end{table}
%

The performance of this policy is very similar to round robin policy
In both these policies, in some fraction of auctions,  selected advertisers are not scheduled because their valuation is lower than the reserve price.

It is possible that one can modify these policies in the following manner. The set of advertisers is selected in auction based on their valuation, i.e., if their valuation is higher than reserve price. Later, round robin or randomized policy can be used to select and schedule one advertiser for each round of auction. Now, it can be expected that the DSP wins more number of impressions for advertisers. This increases the expected revenue of both the seller and the advertisers. 


\subsection{Impression constraint advertisers at DSP}

Using Lagrangian relaxation approach for 
problem~\eqref{eq:opt-prob}, we obtain
\begin{eqnarray*}
L(x,\lambda) = \sum_{t=1}^{T} \sum_{n = 1}^{N} (v_{n,t} -q) x_{n,t} + \sum_{n = 1}^{N}
\lambda_n \left(\sum_{t=1}^{T} x_{n,t} - \Delta_n\right).
\end{eqnarray*}
Here, $\lambda_n \geq 0$ is Lagrangian variable associated with demand of advertiser $n.$
After simplification, we have 
\begin{eqnarray*}
	L(x,\lambda) = \sum_{t=1}^{T} \sum_{n = 1}^{N} (v_{n,t} + \lambda_n  -q) x_{n,t} - \sum_{n = 1}^{N}
	\lambda_n  \Delta_n.
\end{eqnarray*}
We represent the additional constraint as follows.
\begin{eqnarray*}
	C = \left\{ x ~ \bigg\vert~  \sum_{n = 1}^{N} {x}_{n,t} \leq  1, 1 \leq t \leq T;  \right. \\ \left. 
	{x}_{n,t} \in \{0,1 \} \ \ \mbox{for $n\in \mathcal{N},$ $1 \leq t \leq T$} \right\}.
\end{eqnarray*}
We have following inequality $\mathrm{Opt}(P1) \leq \max_{x \in C} L(x, \lambda)$ and 
Let 
\begin{eqnarray*}
h(\lambda) &=& \max_{x \in C} L(x, \lambda) \\
&=& \max_{x \in C} \left[  \sum_{t=1}^{T} \sum_{n = 1}^{N} (v_{n,t} + \lambda_n  -q) x_{n,t} \right] - \sum_{n = 1}^{N} 
\lambda_n  \Delta_n.
\end{eqnarray*}
Moreover DSP may schedule $n$th advertiser if $v_{n,t} + \lambda_n - q \geq 0,$ otherwise DSP does not that advertiser. This is because, for not scheduling, the  payoff is zero instead of negative. This gives us an intuition that, higher demand of $n$th advertiser may correspond to higher values of $\lambda_n.$ This boosts the valuation to $(v_{n,t} + \lambda_n).$ This introduces aggressive bidding behavior of advertisers with high demand of impressions and hence these advertisers are  scheduled more frequently than others.

Note that this is a linear program with integer constraints. Thus, it is difficult to solve. 
An approach is to further relax the integer constraints and consider $0 \leq x_{n,t} \leq 1.$ Introduction of this relaxation increases the size of constraints.  
Hence, we can have 
\begin{eqnarray*}
	\widetilde{C} = \left\{ x ~ \bigg\vert~  \sum_{n = 1}^{N} {x}_{n,t} \leq  1, 1 \leq t \leq T;  \right. \\ \left. 
 0 \leq 	{x}_{n,t} \leq 1 \ \ \mbox{for $n\in \mathcal{N},$ $1 \leq t \leq T$}, \right\}.
\end{eqnarray*}
and $C \subset \widetilde{C}.$
Then relaxed Lagrangian problem is 

\begin{eqnarray*}
	h^R(\lambda) &=& \max_{x \in \widetilde{C}} L(x, \lambda) \\
	&=& \max_{x \in \widetilde{C} } \left[  \sum_{t=1}^{T} \sum_{n = 1}^{N} (v_{n,t} + \lambda_n  -q) x_{n,t} \right] - \sum_{n = 1}^{N} 
	\lambda_n  \Delta_n.
\end{eqnarray*}
Also, $h(\lambda) \leq  h^R(\lambda).$ Then, Lagrangian dual of the relaxation of problem~\eqref{eq:opt-prob} is given by 
\begin{equation*} 
h^{R}_{LD} =\min_{\lambda \geq 0} h^R(\lambda).
\end{equation*} 
Note that $h^R(\lambda)$ is piece-wise linear and convex in $\lambda.$ Then subgradient algorithm for $\lambda$ is as follows. 
\begin{equation*}
\lambda_n^{k+1} = \max \left\{ 
0, \lambda_n^{k} + s^{k} \left(\sum_{t=1}^{T} x_{n,t}(\lambda^k) - \Delta_n\right)
 \right\},
\end{equation*}
for $n=1,2,\cdots, N.$ Here, $s^k$ is step sizes. The convergence of subgradient algorithm is obtained for suitably  selected step sizes. More detail on step size selection is given in \cite[Chapter $10,$ page no. $502$]{Bertsekas98}.  

\subsubsection{Numerical examples}

\begin{table}[h]
	\begin{center}
		\caption{Advertiser's revenue with fixed valuation $v=2.5,$ $q=1$ and impression demand  }
		\label{table:DSP-val-model-demand-impress}
		\begin{tabular}{l|c|c|}
			\cline{2-3}
			& \multicolumn{2}{c|}{$q=1$}  \\ \hline
			\multicolumn{1}{|l|}{Adv.} & Impressions         & Revenue     \\ \hline
			\multicolumn{1}{|l|}{1}    &485        &  $0.07$    \\ \hline
			\multicolumn{1}{|l|}{2}    & 984         &  $0.15$ \\ \hline
			\multicolumn{1}{|l|}{3}    & 6002         &  $0.91$   \\ \hline
			\multicolumn{1}{|l|}{4}    & 498         &  $0.07$ \\ \hline
			\multicolumn{1}{|l|}{5}    & 2031         &  $0.3$   \\ \hline
		\end{tabular}
	\end{center}
\end{table}

We next present few numerical examples with simple heuristic greedy algorithms. 
In the first example, we consider the valuation of all advertisers to be equal and it is constant for all rounds of auctions. Moreover, this valuation is higher than reserve price $q.$ We use the following parameters. $v = 2.5,$  $q=1,$ $N=5$ and $T=10000$ and demand for impressions $\Delta = [400, 800, 4800,400, 1600].$ Define $\xi_n = \frac{\Delta_n}{\sum_{n=1}^{N} \Delta_n}.$ Then $\xi = [0.05, 0.1, 0.6, 0.05,0.2].$
We use  a greedy algorithm in which an advertiser is selected according to their demand, i.e., advertiser $n$ is selected with prob. $\xi_n.$ Table~\ref{table:DSP-val-model-demand-impress} shows the expected revenue of  seller and the number of impressions for different advertisers. Expected revenue of the seller is $1.$ 
Notice that a simple greedy algorithm at DSP meets the demands of all advertisers.

\begin{table}[h] 
	\begin{center}
		\caption{Advertiser's revenue with fixed valuation according to lognormal distribution function and impression demand  }
		\label{table:DSP-val-model-demand-impress-2}
		\begin{tabular}{l|c|c|}
			\cline{2-3}
			& \multicolumn{2}{c|}{$q=1$}  \\ \hline
			\multicolumn{1}{|l|}{Adv.} & Impressions         & Revenue     \\ \hline
			\multicolumn{1}{|l|}{1}    & 260        &  $0.04$    \\ \hline
			\multicolumn{1}{|l|}{2}    & 519         &  $0.09$ \\ \hline
			\multicolumn{1}{|l|}{3}    & 2981         &  $0.52$   \\ \hline
			\multicolumn{1}{|l|}{4}    & 235         &  $0.04$ \\ \hline
			\multicolumn{1}{|l|}{5}    & 1000         &  $0.17$   \\ \hline
		\end{tabular}
	\end{center}
\end{table}

In second numerical example, the valuation of advertisers is fixed and it is drawn according to lognormal distribution function. In this example, DSP selects advertiser $n$ with prob. $\xi_n$ and it is scheduled  in auction $t$ if $v_{n,t} > q.$ Otherwise no advertiser is scheduled in auction $t.$ The expected revenue of advertisers and number of impressions is given in Table~\ref{table:DSP-val-model-demand-impress-2}. The seller's expected revenue is $0.5.$ Observe that using this policy, the impression demands of some advertisers are not fulfilled. 

\begin{table}[h]
	\begin{center}
		\caption{Advertiser's revenue with fixed valuation according to lognormal distribution function and impression demand: Example-$3$  }
		\label{table:DSP-val-model-demand-impress-3}
		\begin{tabular}{l|c|c|}
			\cline{2-3}
			& \multicolumn{2}{c|}{$q=1$}  \\ \hline
			\multicolumn{1}{|l|}{Adv.} & Impressions         & Revenue     \\ \hline
			\multicolumn{1}{|l|}{1}    & 1024        &  $0.19$    \\ \hline
			\multicolumn{1}{|l|}{2}    & 1607         &  $0.28$ \\ \hline
			\multicolumn{1}{|l|}{3}    & 3749         &  $0.64$   \\ \hline
			\multicolumn{1}{|l|}{4}    & 1036         &  $0.19$ \\ \hline
			\multicolumn{1}{|l|}{5}    & 2317         &  $0.40$   \\ \hline
		\end{tabular}
	\end{center}
\end{table}

In the third example we consider another variant of the preceding algorithm.  Here,  advertisers with valuation above the reserve price are selected. Then, DSP schedules  advertisers according to their demand of impressions. The results are shown in Table~\ref{table:DSP-val-model-demand-impress-3}. Now observe that the  impression demand of all advertisers is fulfilled, except advertiser $3$ which is very higher than other advertisers.  This simple variation performs better than the preceding algorithm. Thus, the expected revenue of advertisers is higher and the  expected revenue of a seller is $0.97.$ 

\begin{table}[h]
	\begin{center}
		\caption{Advertiser's revenue with fixed valuation according to Lognormal distribution function and impression demand: Example-$4$  }
		\label{table:DSP-val-model-demand-impress-4}
		\begin{tabular}{l|c|c|}
			\cline{2-3}
			& \multicolumn{2}{c|}{$q=1$}  \\ \hline
			\multicolumn{1}{|l|}{Adv.} & Impressions         & Revenue     \\ \hline
			\multicolumn{1}{|l|}{1}    & 814        &  $0.1557$    \\ \hline
			\multicolumn{1}{|l|}{2}    & 1177         &  $0.1972$ \\ \hline
			\multicolumn{1}{|l|}{3}    & 5046         &  $0.8235$   \\ \hline
			\multicolumn{1}{|l|}{4}    & 810         &  $0.1534$ \\ \hline
			\multicolumn{1}{|l|}{5}    & 1931         &  $0.3342$   \\ \hline
		\end{tabular}
	\end{center}
\end{table}

In our final numerical example, we use the insight developed from Lagrangian relaxation of the problem. If the demand of advertiser is high, then the valuation of that advertiser is boosted by having higher Lagrangian multiplier. Motivated from this, in this example the valuation of advertiser $3$ is increase marginally for all auctions by adding $\lambda_3 = 0.25$ and for other advertisers $\lambda_n =0,$ $n \neq 3.$  Here, we use the algorithm studied in example 3. This is given in Table~\ref{table:DSP-val-model-demand-impress-4}.  The seller's expected revenue is $0.98.$ Note that using this simple variation, DSP can fulfill  the impression demand of all advertisers.




\end{document}